# Size and Stoichiometric Dependence of Thermal Conductivities of $In_xGa_{1-x}N$: A Molecular Dynamics Study

Bowen Wang[a], Xuefei Yan[a], Hejin Yan[a], Yongqing Cai[a]*

[a]Institute of Applied Physics and Materials Engineering, University of Macau, Macau, China

*Email: yongqingcai@um.edu.mo

**Abstract:** The thermal conductivities $\kappa$ of wurtzite $In_xGa_{1-x}N$ are investigated using equilibrium molecular dynamics (MD) method. The $\kappa$ of $In_xGa_{1-x}N$ rapidly declines from InN ($\kappa_{InN}$ = 141 W/mK) or GaN ($\kappa_{GaN}$ = 500 W/mK) to $In_xGa_{1-x}N$ ($x \neq$ or 1), and reaches a minimum ($\kappa_{min}$ = 19 W/mK) when $x$ is around 0.5 at 300 K. The mean free path (MFP) of $In_xGa_{1-x}N$, ranging from 2 to 5 nm and following the same trend with the $\kappa$, is extrapolated in our simulation and a parabolic relationship between $x$ and MFP is established. We find that the $\kappa$ of $In_xGa_{1-x}N$ decreases with increasing temperatures. The evolution of $\kappa$ of In$_x$Ga$_{1-x}$N is also examined by projecting the momentum-energy relationship of phonons from MD trajectories. The phonon dispersion and phonon density of states for In$_x$Ga$_{1-x}$N reflect a slightly more flattened dispersive phononic curve of the alloying system. Despite an overestimated $\kappa$ than experimental values, our calculated $\kappa$ at 300 K agrees well with the results obtained by solving Boltzmann transport equation and also has the same stoichiometric trend with the experimental data. Our study provides the coherent analysis of the effect of thickness, temperature and stoichiometric content on the thermal transport of In$_x$Ga$_{1-x}$N which is helpful for the thermal management of In$_x$Ga$_{1-x}$N based devices.

## 1. INTRODUCTION

Due to a strong light absorption and adjustable bandgap[1, 2], spanning a broad spectrum from infrared (0.69 eV) for InN to ultraviolet (3.4 eV) for GaN, In$_x$Ga$_{1-x}$N compounds are emerging materials highly promising for applications of photovoltaic[3] and light emitting diodes[4]. Despite decades of development, the applications of In$_x$Ga$_{1-x}$N still suffer from issues, such as thermal droop and efficiency droop[5, 6]. Thermal management plays an important role in the efficiency and the endurance of devices for applications of In$_x$Ga$_{1-x}$N. In contrast to the great attention received by GaN from both experiment and simulation, few related experimental and theoretical studies of In$_x$Ga$_{1-x}$N have been reported. Most of the existing experiments on In$_x$Ga$_{1-x}$N[7, 8] only contain several components of $x$. A coherent and systematic understanding of sample size, components, temperatures and other factors on the thermal properties of In$_x$Ga$_{1-x}$N is missing.

The $\kappa$ of single-crystal GaN at room temperature is found to be 253 ± 8.8 W/mK[9] and $\kappa$ of InN, obtained from measurement of InN ceramics, was 45 W/(m·K)[10]. It was reported that the $\kappa$ of In$_x$Ga$_{1-x}$N alloys substantially decreases with the increase of In contents. Pantha et al. experimentally measured that the $\kappa$ of In$_x$Ga$_{1-x}$N alloys are 8.1, 5.4, 2.7, and 1.05 W/mK for $x$=0.16, 0.22, 0.28, and 0.36, respectively at 300 K[8], which is significantly lower than the $\kappa$ of GaN and InN. Adachi[11] used a simplified model of the alloy-disorder scattering to describe the $\kappa$ as a function of In contents at 300 K and got a good agreement between the model and part experimental data. Bahadir et al[2]. also demonstrated that $\kappa$ of In$_x$Ga$_{1-x}$N ($x$ = 0.07 to 0.20) alloys have drastically reduced from 11 to 4.7 W/mK. The same trend was found by Alexander et al[12] and Hua et al[13] through solving the Boltzmann transport equation. Molecular dynamics (MD) simulations were widely used to study the mechanical behaviour of materials and thermally activated evolution of structures[14-17]. MD simulations were performed on the $\kappa$ of GaN in recent years, such as temperature, dimension, and film width[18, 19], but research on the thermal conductivity of In$_x$Ga$_{1-x}$N yet explores. A complete information spanning the whole stoichiometricity is highly critical for the electronic and energy applications and act as the motivation of our current study.

In this paper, we systematically studied thermal conductivity of $In_xGa_{1-x}N$ applying equilibrium molecular dynamics method to uncover the effects of size, temperature excitation and stoichiometricity. The $\kappa$ of $In_xGa_{1-x}N$ film with varying lengths (ranging from 50 to 550 Å), is studied by using Green-Kubo method. The effects of indium content $x$ (from 0.0 to 1.0) and temperature (from 100 to 500 K) on $\kappa$ of $In_xGa_{1-x}N$ have been systematically studied. Importantly, we establish a quantitative relationship of the mean free path (MFP) of phonons with stoichiometricity which allows a facile analysis of the $\kappa$ for any $In_xGa_{1-x}N$ alloys. To better understand the insight on thermal transport of $In_xGa_{1-x}N$, the phonon dispersion and phonon density of states (PDOS) have also been derived from MD trajectories to support the results.

## 2. METHODOLOGY

The equilibrium molecular dynamics (EMD) method is carried out to investigate the thermal transport properties of $In_xGa_{1-x}N$ in this work. The Green-Kubo method [20, 21] within the framework of the fluctuation-dissipation and linear response theorem[22] is used to compute $\kappa$ and especially eligible for perfect crystal systems like Si that have a very long MFP. The thermal conductivity in a particular direction ($\kappa_x$) is related to heat current auto correlation function (HCACF) via the Green-Kubo expression.

$$\kappa_x = \frac{1}{Vk_B T^2} \int_0^\tau \langle j_x(t) \cdot j_x(0) dt \rangle, \qquad (1)$$

where the $V$ is the system volume, $k_B$ is the Boltzmann constant, $T$ is the system temperature, and the angular brackets denote an ensemble average of the auto-correlation of the heat flux $j$. For a system of $N$ atoms described by a general many-body potential with the total potential energy $U$,

$$U = \sum_{i=1}^{N} U_{i(\{r_{ij}\}i \neq j)}, \qquad (2)$$

and the heat current $j$ defined as follows,

$$j = \sum_i \sum_{i \neq j} r_{ij} \frac{\partial U_j}{\partial r_{ij}} \cdot v_i, \qquad (3)$$

where $r_{ij}=r_i-r_j$ and $U_i, r_i$, and $v_i$ are respectively the potential energy, position, and velocity of atom $i$.

The equilibrium $In_xGa_{1-x}N$ has a wurtzite hexagonal crystal structure with three orthogonal directions [0001], [$\bar{1}$100], and [11$\bar{2}$0] which are the $x$, $y$ and $z$ axes as shown in Fig. 1. We performed a series of study to determine the [0001] thermal conductivities of a wurtzite $In_xGa_{1-x}N$ constructed in configuration of film with varying length. The length of our simulated model ranges from 50 to 550 Å. Based our test, the width and thickness of film are kept constant at 50 and 80 Å, which is big enough to resolve the Brillion zone and keep the thermal conductivity forecast from diverging. The indium atoms are randomly distributed in positions equivalent to gallium atoms in $In_xGa_{1-x}N$ and the structure of $In_xGa_{1-x}N$ is built by replacing Ga to In randomly according to the proportion of In content. This work is performed by Large-scale Atomic/Molecular Massively Parallel Simulator[23](LAMMPS). A Stillinger-Weber (SW) potential for $In_xGa_{1-x}N$ created by Zhou et al[24] is used to describe the interactions between Ga, In and N in $In_xGa_{1-x}N$. This SW potential reproduces the experimental data of the equilibrium wurtzite $In_xGa_{1-x}N$ and is used in the MD simulations of $In_xGa_{1-x}N$[25]. By using this SW potential we get the phonon dispersion of GaN which matches well with the reported results[26]. Periodic boundary conditions (PBC) are applied in the $x$, $y$ and $z$ directions and Velocity Verlet algorithm is applied to integrate Newton's equation of motion with a set time step of 1 fs. The simulation process is divided into two parts. In the first part the constant volume and constant temperature (NVT) ensemble with Nosé-Hoover heat bash is applied to equilibrate the simulation box at 300 K. After the system reaches equilibrium and the desired temperature (300 K), the constant volume and constant energy (NVE) is applied to the system for a long time based on the total atoms number until a reasonable decay of HCACF converged $\kappa$ is got. It was reported[26, 27] that significant statistical errors exist among the calculation of thermal conductivity by using molecular dynamics. To reduce statistical errors, our MD simulations take an average time longer than 3 ns and five independent simulations are performed for each specific composition. For the 300 K temperatures, the thermal conductivities in the three coordinate directions [0001], [$\bar{1}$100], and [11$\bar{2}$0] of In$_{0.5}$Ga$_{0.5}$N are presented in Fig. S1 of the supplementary information. With a simulation for 3000 ps, the $\kappa$ is converged and roughly 17 W/mK m in the [0001] direction, 19 W/mK m in the [11$\bar{2}$0] direction, and 17 W/mK in the [$\bar{1}$100] direction. To minimize the statistical error in MD simulations, we generate different random structures of In$_x$Ga$_{1-x}$N for each $x$ and find that the $\kappa$ in In$_x$Ga$_{1-x}$N does not vary much along different directions which is also observed in GaN[27].

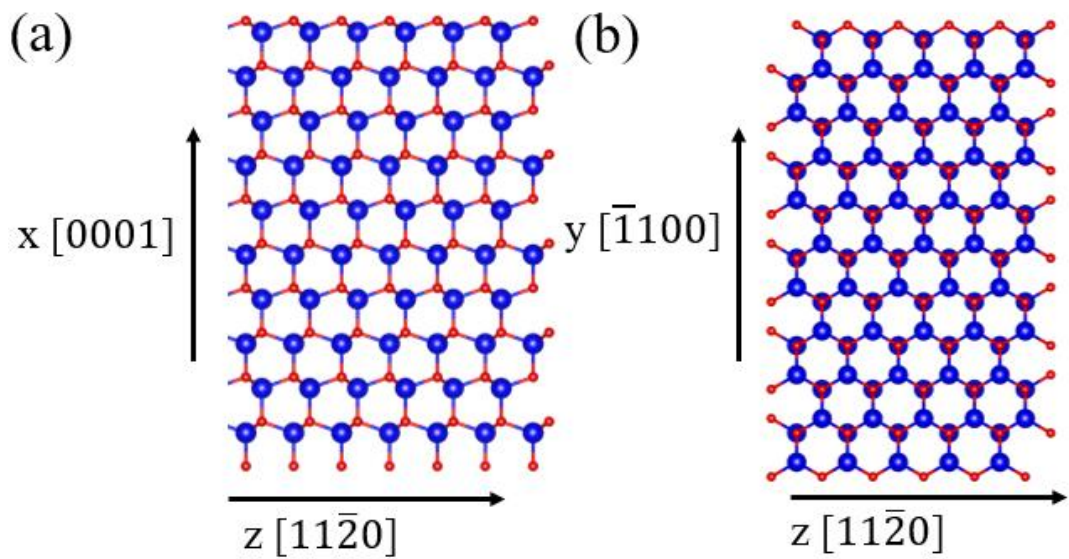

**Figure 1**. Top views from (a) y ($[\bar{1}100]$) and (b) $x$ ([0001]) directions of $In_xGa_{1-x}N$. Red and blue balls represent Ga and N, respectively.

The phonon dispersion and phonon density of states (PDOS) are calculated by DynaPhoPy[28], which extracts quasiparticle phonon frequencies and linewidths from MD trajectories using the normal-mode-decomposition technique[29].

## 3. RESULTS AND DISCUSSION

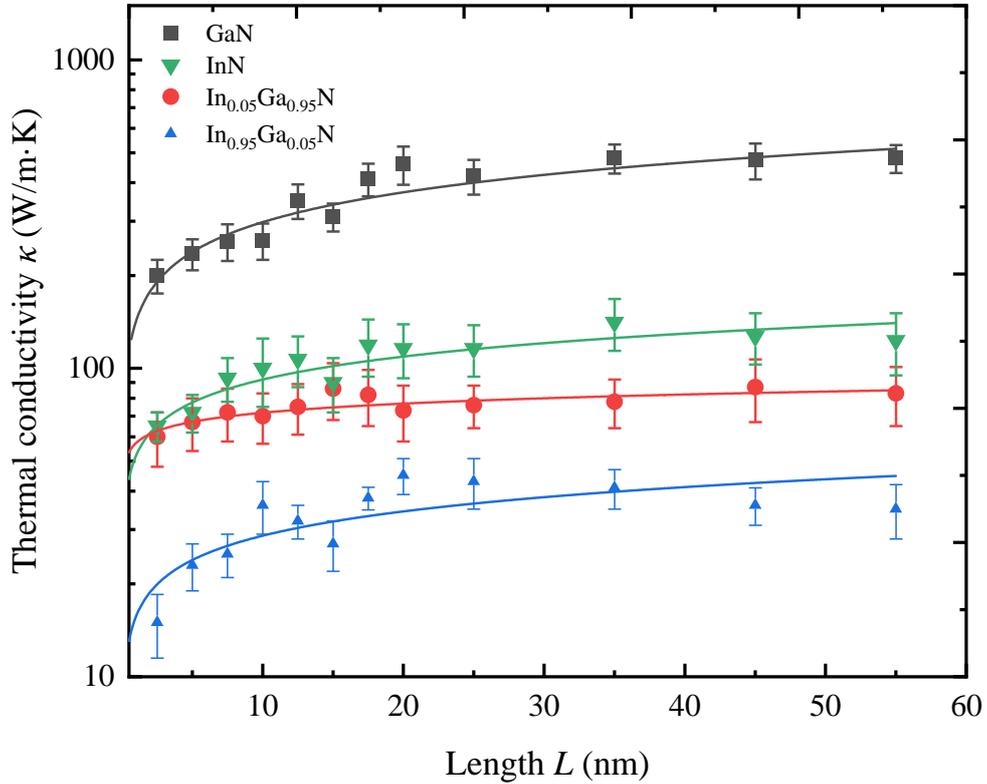

**Figure 2**. Thermal conductivity of $In_xGa_{1-x}N$ ($x$ = 0.1, 0.05, 0.95, 1.0) as a function of length $L$.

As we can see from the Fig. 2, the "$\kappa$" of GaN, InN and the slightly doped

In$_x$Ga$_{1-x}$N ($x$ = 0.05 or 0.95) strongly depend and increase with $L$ (from 10 to 55 nm) obviously. This behavior is similar to some low dimensional nanomaterials in which $\kappa$ increases with the increase of sample length[30]. Moreover, as the length increases, thermal conductivity becomes less sensitive to the variation of length. This is due the fact that when length increases the MFP reduces as phonon-boundary scattering becomes less dominant. The $\kappa$ at small scale can be thought to depend on the phonon MFP, but the $\kappa$ levels off with the further increases of the length. Additionally, we can see the $\kappa$ of slightly doped In$_x$Ga$_{1-x}$N is several times lower than those of pure GaN or InN, which is consistent with the previous studies on the In$_x$Ga$_{1-x}$N[8, 26]. To get more information about the stoichiometric dependence for $\kappa$ of In$_x$Ga$_{1-x}$N, we examined more cases with In content $x$ ranging from 0.1 to 0.9 to calculate the $\kappa$ as a function of length for different stoichiometric series (see Fig. S2 of the supplementary information). The $\kappa$ of In$_x$Ga$_{1-x}$N not always decreases with increasing In content because the converged $\kappa$ values of In$_{0.7}$Ga$_{0.3}$N and In$_{0.3}$Ga$_{0.7}$N are greater than In$_{0.5}$Ga$_{0.5}$N.

According to the kinetic theory of phonon transport[31] and previous studies[26, 32, 33] of the $\kappa$, the relationship between the reciprocal of $\kappa$ and the reciprocal of length scale $L$ follows a linear scaling law[34] as follows:

$$\frac{1}{\kappa} = \frac{1}{\kappa_0}\left(1 + \frac{\lambda}{L}\right), \qquad (4)$$

where the $\kappa$ is the size dependent thermal conductivity, $\kappa_0$ is the converged $\kappa$ with infinite-size sample film, $\lambda$ is the phonon MFP and $L$ is the size of the sample film. In Fig. 3, we show the reciprocal of the $\kappa$ of $In_xGa_{1-x}N$ versus the reciprocal of the length $L$. We fit the data points using a linear function mentioned in Eq. 4 and use the intercept and slope to extrapolate the $\kappa_0$ of infinite sample and MFP of different $In_xGa_{1-x}N$ materials. The converged value of $\kappa_0$ of $In_xGa_{1-x}N$ reaches the lowest value when $x$ is approximately equal to 0.5 and then increases with In content. This result is similar to the reported study on other nitrides alloys[35-39]. The different $\kappa_0$ of $In_{0.9}Ga_{0.1}N$ (< 30 W/m·K) and InN (> 100 W/m·K) in our simulation indicates that there exists an abrupt change between the cases of $In_{0.9}Ga_{0.1}N$ and InN. Due to the difference in the mass (Ga ~ 69.7 and In ~ 114.8 amu) and the strength of N-In and N-Ga bonds[35, 37], high-energy phonons that make a contribution to heat transport are strongly scattered by Rayleigh scattering, leading to the abrupt reduction from binary InN and GaN to ternary $In_xGa_{1-x}N$.

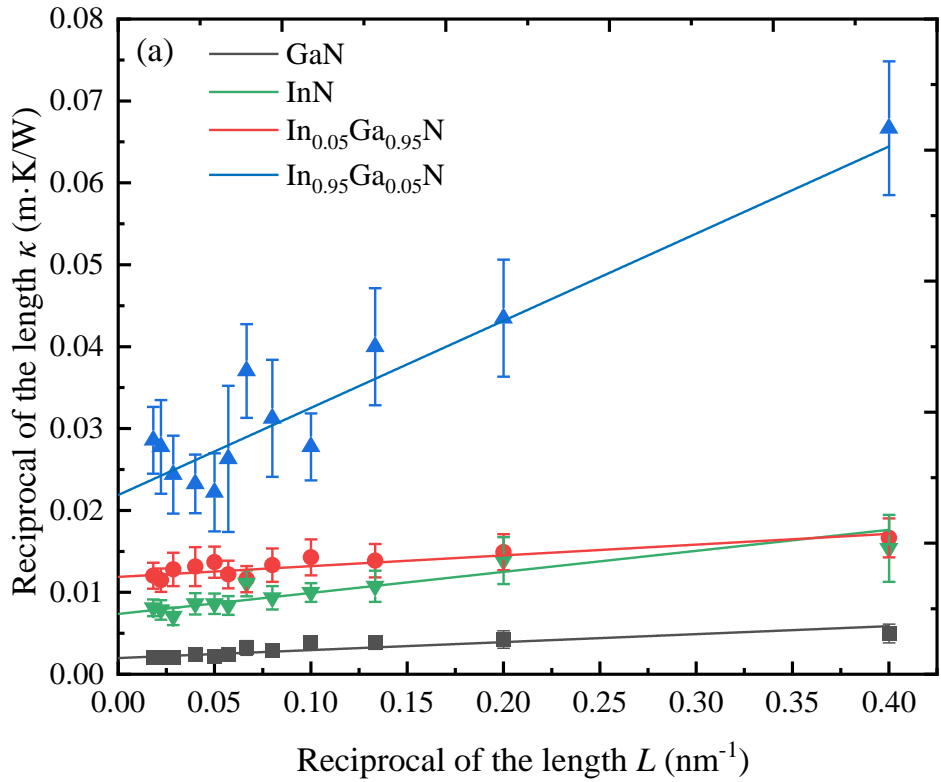

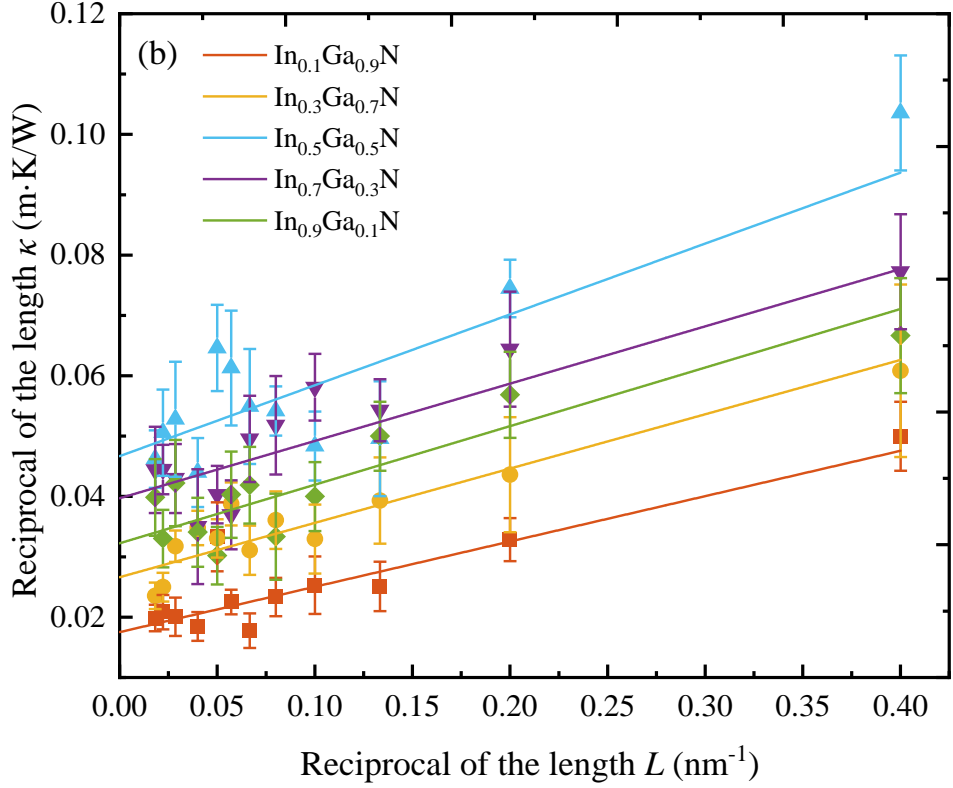

**Figure 3**. Thermal conductivity of $In_xGa_{1-x}N$ as a function of length plotted as $\kappa^{-1}$ versus $L^{-1}$. (a) GaN, InN, and $In_xGa_{1-x}N$ ($x = 0.05, 0.95$) (b) $In_xGa_{1-x}N$ ($x = 0.1$-$0.9$)

The MFPs of $In_xGa_{1-x}N$ at 300 K are also extrapolated and fitted in Fig. 4 to illustrate its dependence on the In content. Compared with pure GaN and InN, the MFP shortens with the modulating the stoichiometry ($x$) after introducing the additional substituting component. A minimum value is achieved when the $x$ is approximately equal to 0.6 and the reduction of MFP may be responsible for the reduction of $\kappa$ of $In_xGa_{1-x}N$. We have also fit this variation and the following relationship between MFP and $x$ is obtained:

$$MFP_{(In_xGa_{1-x}N)} = MFP_{(GaN)} - 8x(1-0.8x),$$

where the $MFP_{(In_xGa_{1-x}N)}$ is the mean free path of $In_xGa_{1-x}N$, $MFP_{(GaN)}$ (5 nm) is the mean free path of GaN, $x$ is the In content.

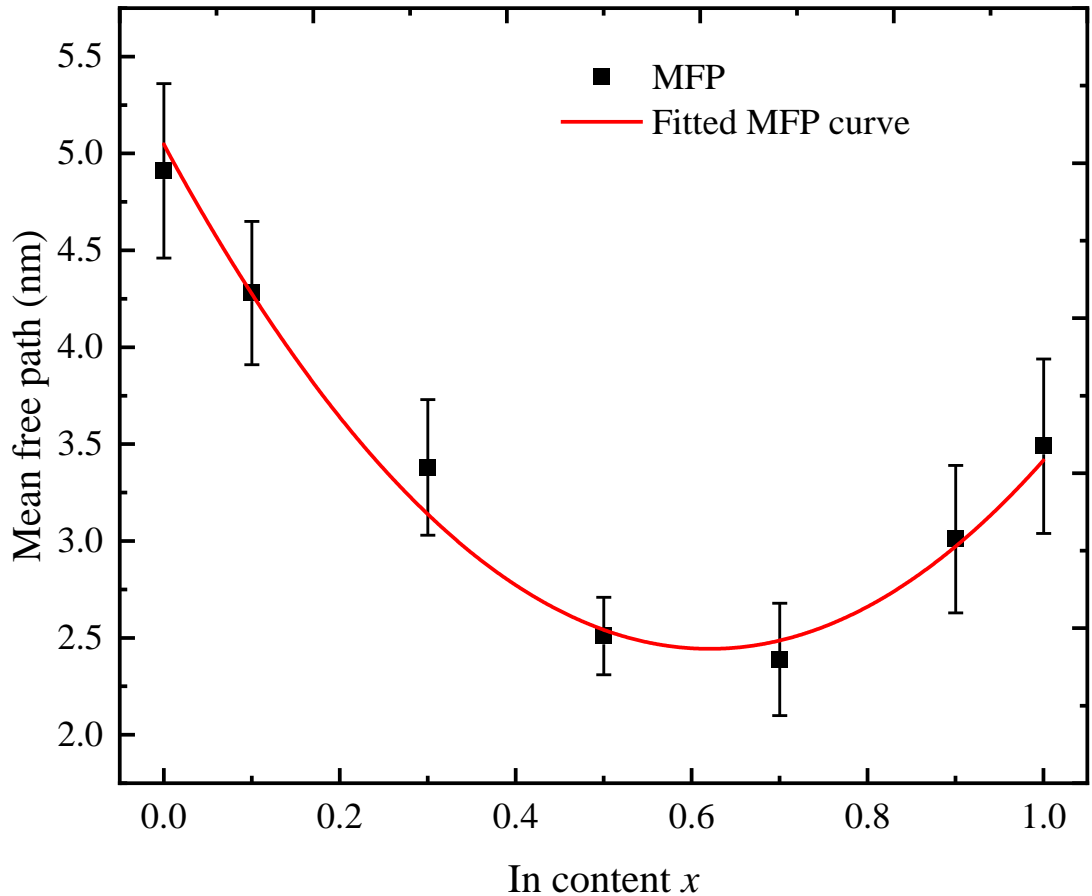

**Figure 4.** The extrapolated mean free path of $In_xGa_{1-x}N$ (from $x = 0$ to 1).

Fig. 5 shows the relation of $\kappa$ as a function of In components at five different temperatures (100 to 500 K). The $\kappa$ of $In_xGa_{1-x}N$ decreases with increasing temperatures as expected for a phonon-dominated crystalline material. The slope reduces and $\kappa$ levels off with increasing temperatures. At high temperatures, i.e., Debye temperature, phonon-phonon Umklapp scattering becomes dominant as the number of high-frequency phonons grows, resulting in a drop in the phonon-phonon scattering coefficient. Due to the increase of phonon density and a shorter MFP, the $\kappa$ of $In_xGa_{1-x}N$ decrease with the increasing In content. The lowest $\kappa$ is achieved for $In_xGa_{1-x}N$ with $x$ equal to 0.5 resulting in value of 19 W/mK, which is much lower in comparison to those of GaN and InN at 300 K.

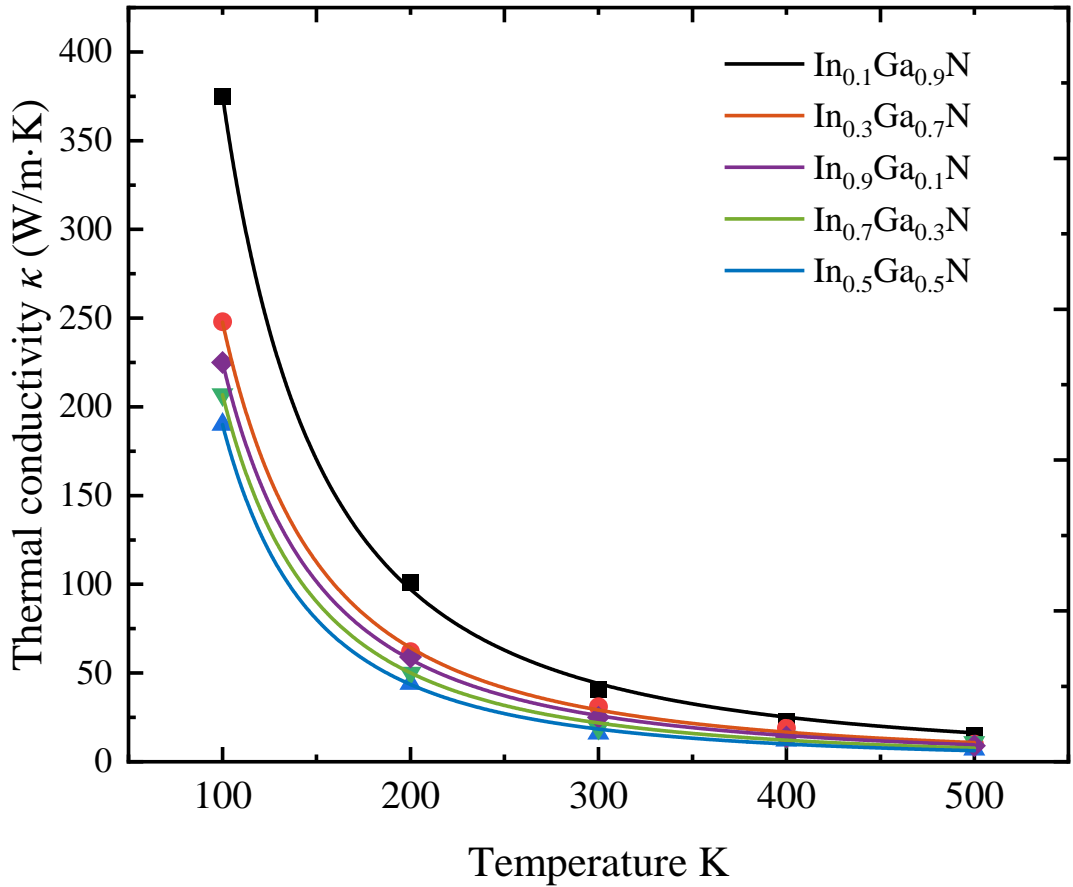

**Figure 5.** $\kappa$ of $In_xGa_{1-x}N$ film as a function of In contents at different temperatures from 100 to 500 K.

We note that some previous experimental and theoretical studies investigated on the $\kappa$ of $In_xGa_{1-x}N$, but only spanning a narrow region of $x$. Herein, in this work a complete spectrum of $x$ is covered. Fig. 6 shows the $\kappa$ of the $In_xGa_{1-x}N$ alloy as a function of the In content fraction $x$, compared with other studies mentioned above. It can be seen that our results have the same trend of dependence of In contents with the Adachi's and Tong's results. The measured $\kappa$ of the alloy, except for pure GaN and InN, is in reasonably good agreement with other studies.

The thermal conductivity $\kappa$ decreases with the increase ratio of In for small $x$, as demonstrated in both the theoretical curves and experiment data. There is a characteristic abrupt reduction when $x$ increases, followed by a gradual reduction to a minimum (nearly at $x = 0.5$). When the In content is larger than 0.9, the $\kappa$ starts to increase rapidly, approaching the bulk InN crystal value. It is obvious that all three curves show reduced $\kappa$ of $In_xGa_{1-x}N$ compared with GaN and InN. Similar trend is also observed in $Al_xGa_{1-x}N$[8, 40], which was attributed to the scattering of phonons due to the alloy disorder[40-43].

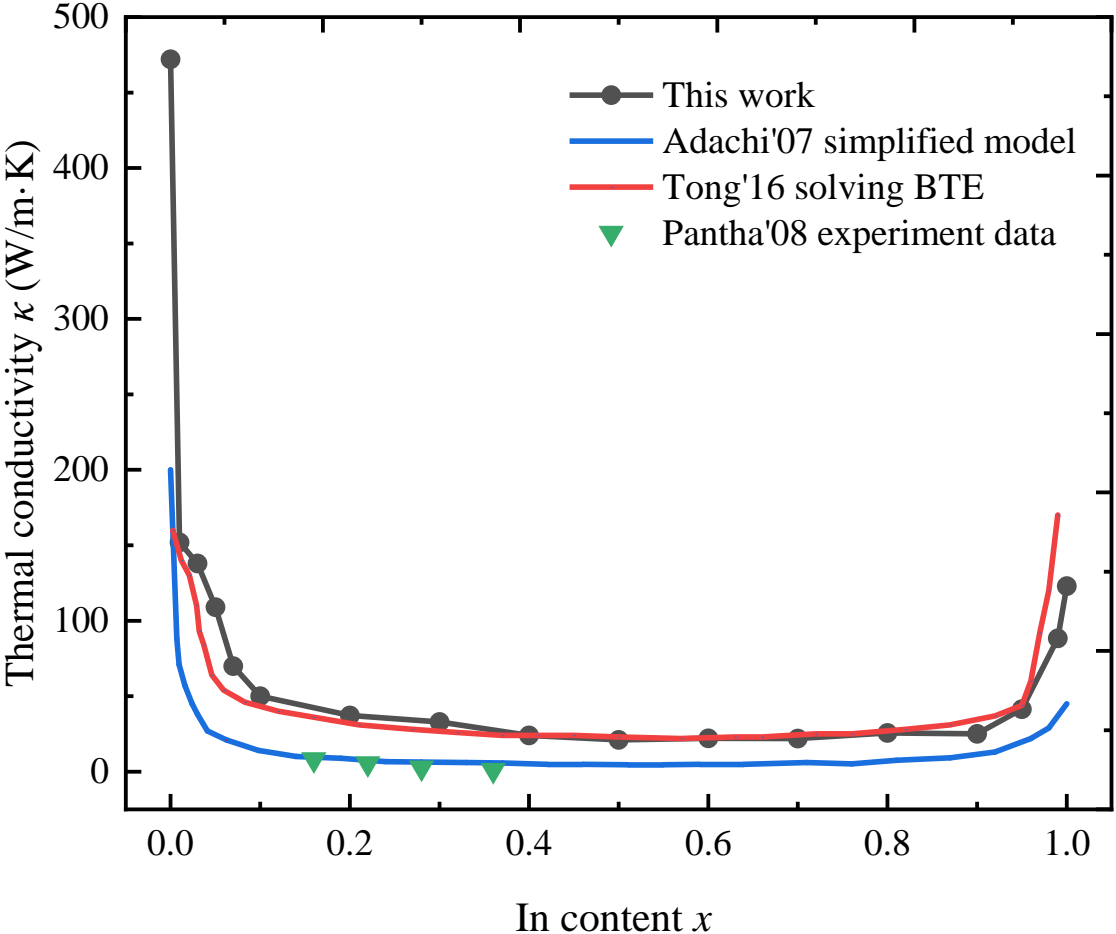

**Figure 6**. The $\kappa$ of $In_xGa_{1-x}N$ as a function of In content $x$ at room temperature.

The black line corresponds to this work (EMD method), red line corresponds to simplified model of alloy-disorder scattering[11], blue line corresponds to solving BTE result[13], green points correspond to experiment data[8].

It is worth mentioning that our predicted $\kappa$ of GaN and InN are 500 and 141 W/mK at room temperature respectively which are higher than the experiment values, while those values of In$_x$Ga$_{1-x}$N ($x$ not equal to 0 or 1) fits well with the BTE results. This is because the potential from Zhou et al[24] is a ternary empirical potential and it might not describe the binary alloy system well. By using the binary SW potential for GaN[44], Zhou et al[18, 27] already got results agree with the experiment data of GaN. Moreover, it is reasonable if we take the following reasons into account. (i) In the simulation, the ideal alloying models of In$_x$Ga$_{1-x}$N are usually used. Due to the lattice mismatch of In$_x$Ga$_{1-x}$N[45] (the mismatch of lattice constant between GaN and InN is 10.7% and 15.0% for $a$ and $c$ constants respectively) and uncertainties related with materials synthesis, the crystal used in the experiment usually contains many pre-exist defects which cause reduction of $\kappa$. Moreover, it was been reported that the localization of phonons around the defects leading the reduction[10, 46, 47] of $\kappa$. (ii) MD calculation using empirical potentials is an approximation of the quantum mechanical calculation. The selection of potential energy in MD simulation may only be used to predict the material trend, and there are usually some deviations in the numerical value.

To further understand the reason for the reduction of $\kappa$ with addition of In content, the phonon dispersion and phonon density of states of GaN, InN and In$_{0.5}$Ga$_{0.5}$N are calculated at 300 K, as shown in Fig. 7. As can be seen from Fig. 7 (a) - (c), there is a large bandgap between low and high frequency (333 ~ 670 $cm^{-1}$) branches. These results agree with the result of GaN found by Qin et al [48]. Furthermore, the low-frequency branches exhibit significant redshifts of about $85 cm^{-1}$ as the In content increases. The main reason for such a change in the phonon dispersion is the different atomic masses of Ga and In ($m_{In} > m_{Ga}$). Then, comparing the corresponding PDOS in Fig. 7(d), we further confirm that the higher magnitude of the peak of In$_{0.5}$Ga$_{0.5}$N for high-frequency modes is responsible for flatter phonon dispersion curve which further indicates the lower phonon group velocity and hence the reduction of $\kappa$ in the In$_x$Ga$_{1-x}$N.

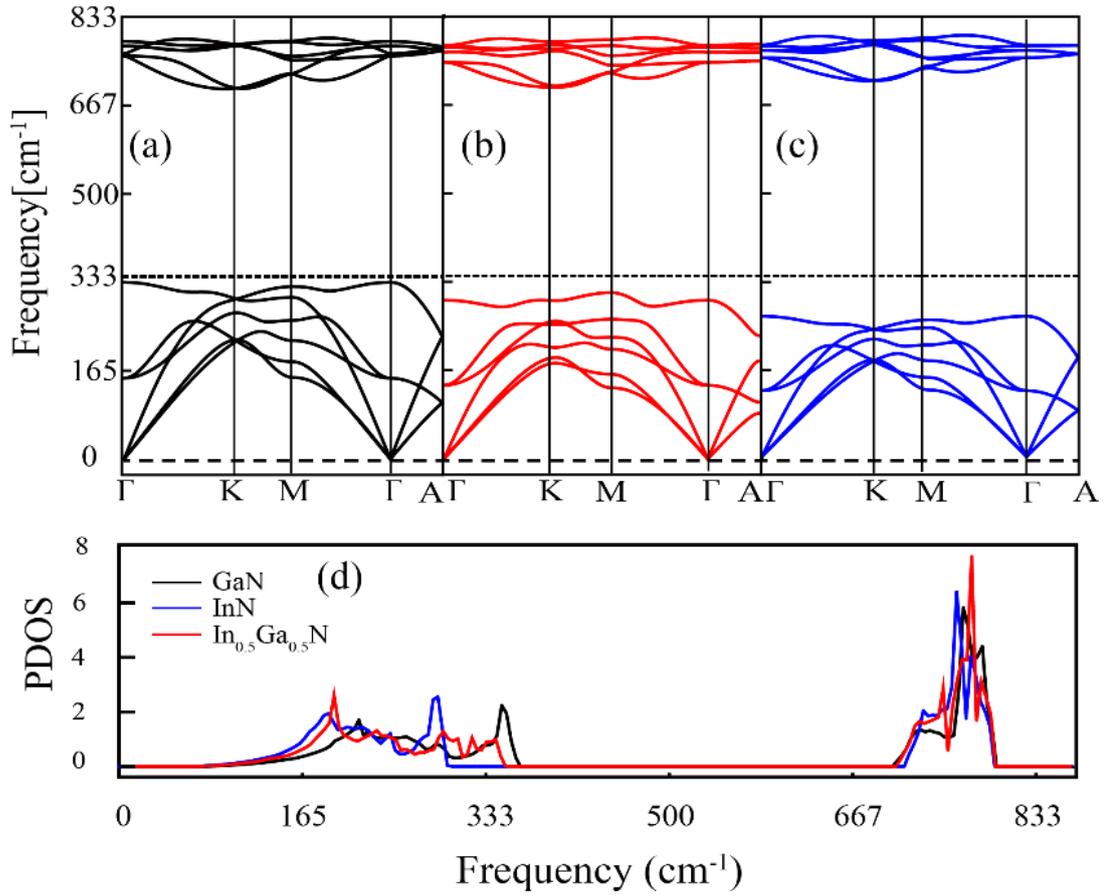

**Figure 7**. Phonon dispersion and PDOS of (a) GaN, (b) In$_{0.5}$Ga$_{0.5}$N, (c) InN calculated using the SW empirical potential approaches. (d) Compiled phonon density of states PDOS for comparison.

## 4. SUMMARY

In conclusion, EMD method is applied to explore the thermal transport properties of wurtzite In$_x$Ga$_{1-x}$N film. The variation of thermal conductivities with sample size and temperatures are reported in our simulation. Compared with the GaN and InN, alloying In$_x$Ga$_{1-x}$N ($x$ not equal to 0 or 1) has a lower MFP due to atomic mixture and alloying and associated strain gradient and local lattice distortion. Similar to the GaN, the thermal conductivities of $In_xGa_{1-x}N$ decrease with the increasing temperature and we found the $\kappa$ of $In_xGa_{1-x}N$ is also sensitive to the variation of length. This phenomenon in agreement with the reported GaN studies and the BTE prediction of $In_xGa_{1-x}N$. The effect of components on the $\kappa$ of $In_xGa_{1-x}N$ is also studied in our simulation. Our results show that there exists a significant reduction of $\kappa$ of $In_xGa_{1-x}N$, which is consistent with the BTE results and experiment results. Further calculations of the phonon dispersion and PDOS show that the higher magnitude of the peak PDOS is responsible for the reduction of $\kappa$. We have derived the quantitative relationship of the mean free path of phonon with the doping content of In which will be highly useful for the thermal management based on the $In_xGa_{1-x}N$ alloys.

## CRediT authorship contribution statement

Bowen Wang: Writing – original draft, Methodology, Software. Xuefei Yan: Formal analysis, Writing – review & editing. Hejin Yan: Software, Writing – review. Yongqing Cai: Conceptualization, Supervision, Writing – review & editing, Funding acquisition.

## SUPPLEMENTARY MATERIAL

See supplementary material for thermal conductivity of $In_{0.5}Ga_{0.5}N$ along the three coordinate directions as a function of relaxation time at 300 K and the thermal conductivity $\kappa$ of $In_xGa_{1-x}N$ ($x = 0.1, 0.3, 0.5, 0.7, 0.9$) as a function of length.

## ACKNOWLEDGMENT

This work was supported by the University of Macau (SRG2019-00179-IAPME) and the Science and Technology Development Fund from Macau SAR(FDCT-0163/2019/A3), the Natural Science Foundation of China (Grant 22022309) and Natural Science Foundation of Guangdong Province, China (2021A1515010024). This work was performed in part at the High-Performance Computing Cluster (HPCC) which is supported by Information and Communication Technology Office (ICTO) of the University of Macau.

## DATA AVAILABILITY

The data that support the findings of this study are available from the corresponding author upon reasonable request.

## CONFLICT OF INTEREST

The authors have no conflicts to disclose.